\documentclass[pra]{revtex4}%
\usepackage{epsfig}
\usepackage{amsmath}
\usepackage{amsfonts}
\usepackage{amssymb}
\usepackage{bm}
\usepackage{euscript}
\usepackage{times}
\normalfont
\def\beq{\begin{equation}}
\def\eeq{\end{equation}}
\def\0{\otimes}
\def\1{\mbox{1\hskip-.25em l}}
\def\6{\langle}
\def\9{\rangle}
\def\half{\mbox{$1\over2$}}

\def\tr{{\rm tr}\,}
\def\Eq{Eq.~(\ref}

\def\bJ{{\bf J}}

\def\bP{{\bf P}}
\def\bS{{\bf S}}

\def\bep{\mbox{\boldmath $\epsilon$}}
\def\bal{\mbox{\boldmath $\alpha$}}

\def\boa{{\bf a}}
\def\bb{{\bf b}}
\def\be{{\bf e}}
\def\bk{{\bf k}}
\def\bn{{\bf n}}
\def\bp{{\bf p}}
\def\bq{{\bf q}}

\def\bv{{\bf v}}

\def\bx{{\bf x}}
\def\by{{\bf y}}
\def\bz{{\bf z}}

\def\hbk{\hat{\bp}}
\def\hbp{\hat{\bp}}
\def\hbq{\hat{\bq}}
\def\hbx{\hat{\bx}}
\def\hby{\hat{\by}}
\def\hbz{\hat{\bz}}

\def\bP{{\bf P}}
\def\bS{{\bf S}}

\def\mC {{\mathbb C}}
\def\mR {{\mathbb R}}

\def\eR{\EuScript{R}}

\def\eO{\EuScript{O}}

\def\cH{{\cal H}}

\def\cO{{\cal O}}

\def\eR{\EuScript{R}}

\def\eO{\EuScript{O}}

\def\bea{\begin{eqnarray}}
\def\eea{\end{eqnarray}}
\def\ba{\begin{array}}
\def\ea{\end{array}}
\def\bay{\begin{array}}
\def\eay{\end{array}}

\begin{document}

\title{Introduction to relativistic  quantum information}

\author{Daniel R. Terno}\email{dterno@perimeterinstitute.ca}
\affiliation{Perimeter Institute, 31 Caroline St, Waterloo, Ontario, Canada N2L 2Y5}

\begin{abstract}
I discuss the role that relativistic considerations play in
quantum information processing. First I describe how the causality
requirements limit possible multi-partite measurements. Then the
 Lorentz transformations of quantum states are introduced, and
 their implications on physical qubits are described. This is used
 to describe relativistic effects in communication and
 entanglement.
\end{abstract}
\maketitle

\begin{center}
\textit{To the memory of Asher Peres,
teacher and friend}\\
\end{center}
\section{Introduction}

 Information and physics are closely and fascinatingly
intertwined. Their relations become even more interesting when we
leave a non-relativistic quantum mechanics for more exiting
venues. My notes are planned as a guided tour for the first steps
along that road, with  open questions and more involved mergers
left to the remarks and to the last section.

I start from a brief introduction to causality restrictions on the
distributed measurements: the limitations that are imposed by
final propagation velocity of the physical interactions. It is
followed by  the relativistic transformations of the states of
massive particles and photons, from which we can deduce what
happens to qubits which are realized as the discrete degrees of
freedom. Building on this, I discuss the distinguishability of
quantum signals, and briefly touch communication channels and the
bipartite entanglement.

I do not follow a historical order or give all of the original
references. A  review \cite{rmp} is used as the standard reference
on quantum information and relativity. The results of the ``usual"
quantum information are given without any reference: all of them
can be found in at least one of the sources \cite{nc,keyl,bgl}.
Finally, a word about units:  $\hbar=c=1$  are always assumed.

\section{Causality and distributed measurements}
 Here I present the causality constraints on  quantum measurements.
For simplicity, measurements are considered to be point-like
interventions.  First  recall the standard description of the
measurement and the induced state transformation. Consider a
system in the state $\rho$ that is subject to measurement  that is
described by a positive operator-valued measure (POVM)
$\{E_\mu\}$. The probability of the outcome $\mu$ is
\beq
p_\mu=\tr E_\mu\rho,
\eeq
while the state transformation is given by some completely
positive evolution
\beq
\rho\rightarrow\rho'_\mu=\sum_mA_{\mu m}\rho A_{\mu m}^\dagger/p_{\mu},
\qquad \sum_m A_{\mu m}^\dagger A_{\mu m}=E_\mu.
\eeq
If the outcome is left unknown, the update rule is
 \beq
\rho\rightarrow\rho=\sum_{\mu m}A_{\mu m}\rho A_{\mu m}^\dagger.
 \eeq

Now consider a bipartite state  $\rho_{AB}$. The operations of
Alice and Bob are given by the operators $A_{\mu m}$ and $B_{\nu
n}$, respectively. It is easy to see that if these operators
commute,
\beq
[A_{\mu m}, B_{\nu n}]=0, \label{etcr}
\eeq
then the observation statistics of Bob is independent of Alice's
results and vice versa.  Indeed, the probability that Bob gets a
result $\nu$, irrespective of what Alice found, is
\beq
p_\nu=\sum_\mu\tr\Bigl(\,\sum_{m,n}B_{\nu n}\,A_{\mu m}\,\rho\,
  A^\dagger_{\mu m}\,B^\dagger_{\nu n}\Bigr). \label{pnu}
\eeq

Now make use of \Eq{etcr}) to exchange the positions of $A_{\mu
m}$ and $B_{\nu n}$, and likewise those of $A^\dagger_{\mu m}$ and
$B^\dagger_{\nu n}$, and then we move $A_{\mu m}$ from the first
position to the last one in the product of operators in the traced
parenthesis. Since the elements of a POVM satisfy $\sum_\mu
E_\mu=\1$, \Eq{pnu}) reduces to
\beq
p_\nu=\tr\Bigl(\sum_n B_{\nu n}\,\rho\,B^\dagger_{\nu n}\Bigr),
\eeq
whence all the expressions involving Alice's operators $A_{\mu m}$
have totally disappeared. The statistics of Bob's result are not
affected at all by what Alice may simultaneously do somewhere
else. This proves that Eq.~(\ref{etcr}) indeed is a sufficient
condition for no instantaneous information transfer. In
particular, the local operations $A\otimes\1_B$ and $\1_A\otimes
B$ are of this form.

Note that any classical communication between distant observers
can be considered  a kind of long range interaction.  The
propagation of signals is, of course, bounded by the velocity of
light. As a result, there exists a partial time ordering of the
various interventions in an experiment, which defines the notions
earlier and later. The input parameters of an intervention are
deterministic (or possibly stochastic) functions of the parameters
of earlier interventions, but not of the stochastic outcomes
resulting from later or mutually spacelike interventions
\cite{rmp}.

Even these apparently simple notions lead to non-trivial results.
Consider a separable bipartite superoperator $T$,
\beq
T(\rho)=\sum_k M_k\rho M_k^\dag, \qquad M_k=A_k\otimes B_k,
\eeq
where the operators $A_k$ represent operations of Alice and $B_k$
those of Bob. Not all such superoperators can be implemented by
local transformations and classical communication (LOCC)
\cite{ben99}. This is the foundation of the ``non-locality without
entanglement''.

A classification of bipartite state transformations was introduced
in \cite{bek01}. It consists of the following categories. There
are {\it localizable\/} operations that can be implemented locally
by Alice and Bob, possibly with the help of prearranged ancillas,
but without classical comunication. Ideally, local operations are
instantaneous, and the whole process can be viewed as performed at
a definite time. A final classical output of such distributed
intervention will  be obtained at some point of the (joint) causal
future of
 Alice's and Bob's interventions. For {\it semilocalizable\/} operations, the
requirement of no communication is relaxed and one-way classical
communication is possible. It is obvious that any tensor-product
operation $T_{\rm A}\otimes T_{\rm B}$ is localizable, but it is
not a necessary condition.  For example the Bell measurements,
which distinguishes between the four standard bipartite entangled
qubit states,
\beq |\Psi^\pm\rangle := {1\over \sqrt 2}(|0\rangle |1\rangle \pm
|1\rangle |0\rangle), \qquad
 |\Phi^\pm\rangle := {1\over \sqrt
2}(|0\rangle |0\rangle \pm |1\rangle |1\rangle),
\eeq
are localizable.

 Other classes of bipartite operators are defined
as follows: Bob performs a local operation $T_{\rm B}$ just before
the global operation $T$. If no local operation of Alice can
reveal any information about $T_{\rm B}$, i.e., Bob cannot signal
to Alice, the operation $T$ is {\it semicausal\/}. If the
operation is semicausal in both directions, it is {\it causal\/}.
In many cases it is easier to prove causality than localizability
(see Remark 3). There is a necessary and sufficient condition for
the semicausality (and therefore, the causality) of operations
\cite{bek01}.

These definitions of causal and localizable operators appear
equivalent. It is easily proved that localizable operators are
causal. It was shown that semicausal operators are always
semilocalizable \cite{egg02}. However, there are causal operations
that are not localizable \cite{bek01}.

It is curious that while a complete Bell measurement is causal,
the two-outcome incomplete Bell measurement is not. Indeed,
consider a two-outcome PVM
\beq
E_1=|\Phi^+\9\6\Phi^+|, \qquad E_2=\1-E_1.
\eeq
 If the initial state is
$|01\9_{\rm AB}$, then the outcome that is associated with $E_2$
always occurs and Alice's reduced density matrix after the
measurement is $\rho_{\rm A}=|0\9\6 0|$. On the other hand, if
before the joint measurement Bob performs a unitary operation that
transforms the state into $|00\9_{\rm AB}$, then the two outcomes
are equiprobable, the resulting states after the measurement are
maximally entangled, and Alice's reduced density matrix is
$\rho_{\rm A}=\half\1$. A simple calculation shows that after this
incomplete Bell measurement two input states $|00\9_{\rm AB}$ and
$|01\9_{\rm AB}$  are distinguished by Alice with a probability of
0.75.

Here is another example of a  semicausal and semilocalizable
measurement which can be executed with one-way classical
communication from Alice to Bob.  Consider a PVM measurement,
whose complete orthogonal projectors are
\beq |0\9\0|0\9, \quad
|0\9\0|1\9, \quad |1\9|+\9, \quad |1\9\0|-\9,\label{examp1}
\eeq
where $|\pm\9=(|0\9\pm|1\9)/\sqrt{2}$. The Kraus matrices are
\beq
A_{\mu j}=E_\mu\delta_{j0}\label{orthopvm},
\eeq
From the properties of complete orthogonal measurements
\cite{bek01}, it follows that this operation cannot be performed
without Alice talking to Bob. A protocol to realize this
measurement is the following. Alice measures her qubit in the
basis $\{|0\9, |1\9\}$, and tells her result to Bob. If Alice's
outcome was $|0\9$, Bob measures his qubit in the basis $\{|0\9,
|1\9\}$, and if it was $|1\9$, in the basis $\{|+\9, |-\9\}$.

If one allows for more complicated conditional state evolution
\cite{gr02}, then more measurements are localizable. In particular,
consider a \textit{verification} measurement, i.e.,
 the measurement yields a $\mu$-th result with
certainty, if the state prior to the classical interventions was
given by $\rho=E_\mu$, but without making any specific demand on
the resulting state $\rho'_\mu$.

It is possible to realize  a  verification measurements by means
of a shared entangled ancilla and Bell-type measurements by one of
the parties \cite{vai03}. Verification measurement  of
Eq.~(\ref{examp1}) can illustrate this construction. In addition
to the state to be tested, Alice and Bob share a Bell state
$|\Psi^-\9$. They do not have to coordinate their moves. Alice and
Bob perform tasks independently and convey their results to a
common center, where a final decision is made.

The procedure is based on the teleportation identity
\beq
    |\Psi\rangle_1|\Psi^-\rangle_{23} = \half
  \left(|\Psi^-\rangle_{12} | \Psi\rangle_{3} +
|\Psi^+\rangle_{12} |\tilde \Psi^{(z)}\rangle_{3} +
|\Phi^-\rangle_{12} |\tilde \Psi^{(x)}\rangle_{3}+
|\Phi^+\rangle_{12} |\tilde \Psi^{(y)}\rangle_{3}\right),
\label{ident}
\eeq
where   $|\tilde
\Psi^{(z)}\rangle$ means the state $| \Psi\rangle$
rotated  by $\pi$ around the $z$-axis, etc. The first step of this
measurement corresponds to the first step of a teleportation  of a
state of the spin from $B$ (Bob's site) to $A$ (Alice's site). Bob
and Alice do not perform the full teleportation (which  requires a
classical communication between them). Instead, Bob performs  only
the Bell measurement at his site which leads to one of the
branches of the superposition in the rhs of Eq.~(\ref{ident}).

The second step of the verification measurement is taken by Alice.
Instead of completing the teleportation protocol, she measures the
spin of her particle in the $z$ direction. According to whether
that spin is up or down, she measures the spin of her ancilla in
the $z$ or $x$ direction, respectively. This completes the
measurement and it only remains to combine the local outcomes to
get the result of the nonlocal measurement
\cite{vai03}. This method can be extended to arbitrary Hilbert
space dimensions.

\subsection*{Remarks}
\noindent 1.~ Measurements in quantum field theory are discussed in
\cite{rmp,rm1, rm2}.\medskip

\noindent 2.~ An algebraic field theory approach to statistical independence and to
related topics is presented in \cite{sf}.\medskip

\noindent 3.~ To check the causality of an operation
$T$ whose outcomes are the states $\rho_\mu=T_\mu(\rho)/p_\mu$
with probabilities $p_\mu=\tr T_\mu(\rho)$, $\sum_\mu p_\mu=1$ it
is enough to consider the corresponding superoperator
\beq
T'(\rho):=\sum_\mu  T_\mu(\rho)
\eeq
Indeed, assume that Bob's action prior to the global operation
lead to one of the two different states $\rho_1$ and $\rho_2$.
Then the states $T'(\rho_1)$ and $T'(\rho_2)$ are distinguishable
 if and only if some of the pairs of
states $T_\mu(\rho_1)/p_{\mu1}$ and $T_\mu(\rho_2)/p_{\mu2}$ are
distinguishable. Such probabilistic distinguishability shows that
the operation $T$ is not semicausal.
\medskip

\noindent 3.~ Absence of the superluminal communication makes possible to evade
the theorems on the impossibility of a bit commitment.
  In particular
the protocol RBC2 allows a bit commitment to be indefinitely
maintained with unconditionally security against all classical
attacks, and at least for some finite amount of time against
quantum attacks \cite{k99,k03}. \medskip

\noindent 4.~ In these notes I am not going to deal with the relativistic
localization POVM. Their properties (and difficulties in their
construction) can be found in \cite{rmp}. An exhaustive survey of
the spatial localization of photons is given in
\cite{keller}. Here we only note in passing that if $E(\cO)$ is an
operator that corresponds to the detection of an event in a
spacetime region $\cO$, since they are not thought to be
implemented by physical operations confined to that spacetime
area, the condition $[E(\cO_1),E(\cO_2)]=0$ is not required
\cite{tol, maz}.

\section{Quantum Lorentz transformations}
There is no elementary particle that is called ``qubit". Qubits
are realized by  particular degrees of freedom of more or less
complicated systems. To decide how qubits transform (e.g., under
Lorentz transformations) it may be necessary to consider again the
entire system. In the following our qubit will be either a spin of
a massive particle or a polarization of a photon. A quantum
Lorentz transformation connects the  description of a quantum
state $|\Psi\9$ in two reference frames that are connected by a
Lorentz transformation $\Lambda$ (i.e., their coordinate axes are
rotated with respect to each other and the frames  have a fixed
relative velocity). Then $|\Psi'\9=U(\Lambda)|\Psi\9$, and the
unitary $U(\Lambda)$ is represented  on Fig.~1 below. The purpose
of this section is to explain the elements of this quantum
circuit.

From the mathematical point of view the single-particle states
belong to some irreducible representation of the Poincar\'{e}
group. An  introductory discussion of these representations and
their relations with
 states and quantum fields may be found, e.g., in
\cite{wkt,wei}. Within each
particular irreducible representation there are six commuting
operators. The eigenvalues of two of them are invariants that
label the representation by defining the mass $m$ and the
intrinsic spin $j$. The basis states are labelled by three
components of the momentum $\bp$ and the spin operator $\Sigma_3$.
Hence a generic state is given by
\beq
|\Psi\9=\sum_\sigma\!\int \!d\mu(p)\psi_\sigma(p)|p,\sigma\9.
\eeq
In this formula $d\mu(p)$ is the Lorentz-invariant measure,
\beq
d\mu(p)=\frac{1}{(2\pi)^3}\frac{d^3\bp}{2E(\bp)},
\eeq
where the energy $E(\bp)=p^0=\sqrt{\bp^2+m^2}$. The improper
momentum and spin eigenstates are $\delta$-normalized,
\beq
\6p,\sigma|q,\sigma'\9=(2\pi)^3(2E(\bp))\delta^{(3)}(\bp-\bq)\delta_{\sigma\sigma'},
\eeq
and are complete on the one-particle space, which is
$\cH=\mC^{2j+1}\otimes L^2(\mR^3,d\mu(p))$ for spin-$j$ fields.

\vspace{-6.cm}
\begin{figure}[htbp]
\epsfxsize=0.50\textwidth
\centerline{\epsffile{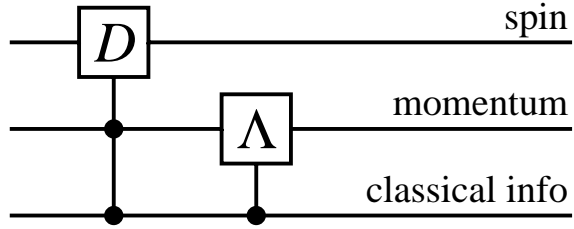}} \vspace*{-6.1cm}
\caption{\small{
Relativistic state transformation as a quantum circuit: the gate
$D$ which represents the matrix $D_{\xi\sigma}[W(\Lambda,p)]$ is
controlled by both the classical information and the momentum $p$,
which is itself subject to the classical information $\Lambda$.}}
\end{figure}

To find the transformation law  we have to be more concrete about
the spin operator. The operator $\Sigma_3(p)$ is a function of the
generators of the Poincar\'{e} group. One popular option is
helicity,
 $\Sigma_3=\bJ\cdot\bP/|\bP|$, which is applicable for both massive and massless particles.
 For massive particles we use the $z$-component of the rest-frame (or
 Wigner spin, that we now describe in the next section.

 \subsection{Massive particles}

The construction involve picking a reference 4-momentum $k$, which
for massive particles is taken to be $k_R=(m,\bf{0})$. The Wigner
spin  $\bS(p)$ is defined to  coincide with the non-relativistic
spin $\bS$ in particle's rest frame. The state of a particle at
rest is labelled  $|k_R,\sigma\9$,
\beq
\bS^2|k_R,\sigma\9=j(j+1)|k_R,\sigma\9,\qquad
\bS_3|k_R,\sigma\9=\sigma|k_R,\sigma\9.
\eeq
The spin states of arbitrary momenta are defined as follows. The
standard rotation-free boost that brings $k_R$ to an arbitrary
momentum $p$, $p^\mu=L(p)^\mu_{~\nu}\/ k^\nu$ is given by
\beq
L(p)=\left(\begin{array}{cccc}
\frac{E}{m} & \frac{p_1}{m} & \frac{p_2}{m} & \frac{p_3}{m}\\
\frac{p_1}{m} & \mbox{\small 1}+ \frac{p_1^2}{m(m+E)} & \frac{p_1 p_2}{m(m+E)} &
\frac{p_1 p_3}{m(m+E)} \\
\frac{p_2}{m} & \frac{p_2 p_1}{m(m+E)} & \mbox{\small 1}+
\frac{p_2^2}{m(m+E)}&\frac{p_2 p_3}{m(m+E)} \\
\frac{p_3}{m} & \frac{p_3 p_1}{m(m+E)} & \frac{p_3 p_2}{m(m+E)} & \mbox{\small 1}+ \frac{p_3^2}{m(m+E)}
\end{array} \label{lp}
\right).
\eeq
The Wigner spin $\bS(p)$ and the one-particle basis states are
defined by
\beq
|p,\sigma\9\equiv U[L(p)]|k_R,\sigma\9,
\qquad\bS_3(p)|p,\sigma\9=\sigma|p,\sigma\9. \label{basis}
\eeq

In deriving the transformation rules we begin with the momentum
eigenstates.  Using the group representation property and
Eqs.~(\ref{basis}) the transformation  is written as
\beq
U(\Lambda)=U[L(\Lambda p)]U[L^{-1}(\Lambda p)\Lambda.
L(p)]U[L^{-1}(p)]
\eeq
The element of the Lorentz group
\beq
W(\Lambda,p)\equiv L^{-1}(\Lambda p) \Lambda L(p), \label{wigw}
\eeq
leaves $k_R$ invariant, $k_R=Wk_R$. Hence it belongs to the
stability subgroup (or Wigner little group) of $k_R$. For
$k_R=(m,\bf{0})$ it is a rotation. Pressing on
\beq
U(\Lambda)|p,\sigma\9=U[L(\Lambda p)]U[W(\Lambda,p)]|k_R,\sigma\9,
\eeq
and as a result,
\beq
U(\Lambda)|p,\sigma\9=\sum_\xi D_{\xi\sigma}[W(\Lambda,p)]|\Lambda
p,\xi\9,
\eeq
where $D_{\xi\sigma}$ are the matrix elements of the
representation of  the Wigner rotation $W(\Lambda,p)$.

We consider only spin-$\half$ particles, so $\sigma=\pm\half$. Any
$2\times2$ unitary matrix can be written as $\hat{D}=\exp(-i\omega
\hat{\mathbf n}\cdot \mbox{\boldmath
$\sigma$})$, where $\omega$ is a rotation angle and $\hat{\mathbf
n}$ is a rotation axis that corresponds to $W(\Lambda,p)$.

The wave functions transform according to $
\psi'_\xi(q)=\6\xi,q|U(\Lambda)|\Psi\9$
so the same state in the Lorentz-transformed frame is
\beq
|\Psi'\9=U(\Lambda)|\Psi\9=\sum_{\sigma,\xi}\int_{-\infty}^\infty\!
D_{\sigma\xi}[W(\Lambda,\Lambda^{-1}p)]\psi_\xi(\Lambda^{-1}p)|\sigma,p\9d\mu(p).\label{state}
\eeq

For pure rotation $\eR$ the three-dimensional (more exactly, 3D
block of 4D matrix; here and in the following we use the same
letter for a 4D and 3D matrix for $\eR\in$SO(3)) Wigner rotation
matrix is the rotation itself,
\beq
W(\eR,p)=\eR,\qquad \forall p=(p^0,\bp).
\eeq
 As a result, the action of
Wigner spin operators on $\cH_1$ is given by than halves of Pauli
matrices that are tensored with the identity of $L^2$.

\subsection{Photons}
The single-photon states are labelled by momentum $\bp$ (the
4-momentum vector is null, $E=p^0=|\bp|$) and helicity
$\sigma_\bp=\pm 1$, so the state with a definite momentum is given
by $\sum_{\sigma=\pm 1}\alpha_\sigma|p,\sigma_\bp\9,$ where
$|\alpha_+|^2+|\alpha_-|^2=1$. Polarization states are also
labelled by 3-vectors $\bep^\sigma_\bp$,
$\bp\cdot\bep_\bp^\sigma=0$, that correspond to the two senses of
polarization of  classical electromagnetic waves. An alternative
labelling of the same state, therefore, is $\sum_{\sigma=\pm
1}\alpha_\sigma|p,\bep^\sigma_\bp\9$.

Helicity is invariant under proper Lorentz transformation, but the
basis states acquire phases.
\
 The little group element $W(\Lambda,p)=L^{-1}(\Lambda p)\Lambda
 L(p)$ is defined with respect to the standard
four-momentum $k_R=(1,0,0,1)$. The standard Lorentz transformation
is
\beq
L(p)=R(\hbk)B_z(u),
\eeq
 where $B_z(u)$ is a pure boost along the $z$-axis with a
velocity $u$ that takes $k_R$ to $(|\bp|,0,0,|\bp|)$ and $R(\hbp)$
is the standard rotation that carries the $z$-axis into the
direction of the unit vector $\hbp$. If $\hbp$ has polar and
azimuthal angles $\theta$ and $\phi$, the standard rotation
$R(\hbp)$ is accomplished  by a rotation by $\theta$ around the
$y$-axis, that is followed by a rotation by $\phi$ around the
$z$-axis. Hence,
\beq
R(\hbp)=\left(\bay{ccccc}
\cos\theta\cos\phi & & -\sin\phi  & & \cos\phi\sin\theta \\
\cos\theta\sin\phi & & \cos\phi & & \sin\phi\sin\theta \\
-\sin\theta & & 0 & & \cos\theta
\eay\right), \label{strot}
\eeq
(here only the non-trivial 3D block is shown).

An arbitrary little group element for a massless particle is
decomposed according to
\beq
W(\Lambda,p)=S(\beta,\gamma)R_z(\xi),
\eeq
where the elements $S(\beta,\gamma)$ form a subgroup that is
isomorphic to the translations of the Euclidean plane and
$R_z(\xi)$ is a rotation around the $z$-axis. We are interested
only in the angle $\xi$, since  $\beta$ and $\gamma$ do not
correspond to the physical degrees of freedom. However, they are
important for gauge transformations. Finally, the little group
elements are represented by
\beq
D_{\sigma'\sigma}=\exp(i\xi\sigma)\delta_{\sigma'\sigma}.\label{dss}
\eeq

It is worthwhile to derive more explicit expressions for $\xi$. I
begin with rotations, $\Lambda=\eR$.  Since rotations form a
subgroup of a Lorentz group, $R^{-1}(\eR\hbp)\eR R(\hbp)$ is a
rotation that leaves $\hbz$ invariant and thus is of the form
$R_z(\omega)$ for some $\omega$. A boost in $(t,z)$ plane and a
rotation around $z$-axis commute,  $[R_z,B_z]=0$, so
\beq
W(\eR,p)=R^{-1}(\eR\hbp)\eR R(\hbp)=R_z(\xi).\label{wr}
\eeq

Any rotation can be described by two angles that give a direction
of the  axis and the third angle that gives the amount of rotation
around that axis. If $\eR\bp=\bq$, we decompose the rotation
matrix as
\beq
\eR=R_{\hbq}(\omega)R(\hbq)R^{-1}(\hbp), \label{omdef}
\eeq
where $R_{\hbq}(\omega)$ characterizes a rotation around $\hbq$,
and $R(\hbq)$ and $R(\hbp)$ are the standard rotations that carry
the $z$-axis to $\hbq$ and $\hbp$, respectively. Using
Eq.~(\ref{wr}) we find that $S=\1$ and the two rotations are of
the same conjugacy class,
\beq
R_z(\xi)=R^{-1}(\eR\hbp)R_{\eR\hbp}(\omega)R(\eR\hbp),\label{rdeco}
\eeq
so we conclude that $\xi=\omega$.

A practical description of polarization states is given by spatial
vectors that correspond to the classical polarization directions.
Taking again $k_R$ as the reference momentum, two basis vectors of
linear polarization are $\bep_{k_R}^1=(1,0,0)$ and
$\bep_{k_R}^2=(0,1,0)$, while to the right and left circular
polarizations correspond $\bep_{k_R}^\pm=(\bep_{k_R}^1\pm
i\bep_{k_R}^2)/\sqrt{2}$.

Phases of the states  obtained by the standard Lorentz
transformations $L(p)$ are set to 1. Since the standard boost
$B_z(u)$ leaves the four-vector $(0,\bep^\pm_{k_R})$ invariant, we
define a polarization basis for any $\bp$ as
\beq
\bep^\pm_\bp=\bep^\pm_{\hbk}\equiv R(\hbk)\bep^\pm_{k_R},\label{conve}
\eeq
while the transformation of polarization vectors under an
arbitrary rotation $\eR$ is given by the rotation itself. To see
the agreement between transformations of spatial vectors and
states, consider a generic state with a momentum $\bp$. Its
polarization is described by the polarization vector
$\bal(\bp)=\alpha_+\bep^+_\bp+\alpha_-\bep^-_\bp$, or by the state
vector $\alpha_+|\bp,+\9+\alpha_-|\bp,-\9$.  Using
Eq.~(\ref{conve}) we see that the transformation of $\bal(\bp)$ is
given by
\beq
\eR\bal(\bp)=R_{\eR\hbp}(\omega)R(\eR\hbk)R^{-1}(\hbk)\bal(\bp)=
R_{\eR\hbk}(\omega)R(\eR\hbk)\bal(k_R)=R_{\eR\hbk}(\omega)\bal(\eR\hbk).
\eeq
 If $\bq=\eR \bp$ the transformation results in
 $\alpha_+e^{i\omega}\bep^+_\bq+\alpha_-e^{-i\omega}\bep^-_\bq$,
and since $\omega=\xi$, it is equivalent to  the state
transformation
\beq
U(\eR)(\alpha_+|p,+\9+\alpha_-|p,-\9)=
\alpha_+e^{i\xi}|q,+\9+\alpha_-e^{-i\xi}|q,-\9.
\eeq

For a general Lorentz transformations the triad
$(\bep^1_\bp,\bep^2_\bp, \hbk)$ is rigidly rotated, but in a more
complicated fashion. To obtain the phase for a general Lorentz
transformation, we decompose the latter into two rotations and a
standard boost $B_z$ along the $z$-axis:
\beq
\Lambda=\eR_2B_z(u)\eR_1.
\eeq
It can be shown that $B_z$ alone does not lead to a phase
rotation. Therefore,
\beq
\xi=\omega_1+\omega_2,
\eeq
where both $\omega_1$ and $\omega_2$ are due to the rotations and
are given by Eq.~(\ref{omdef}). Note that although $B_z(u)$ alone
does not lead to a phase rotation, it can affect the value of
$\omega_2$, since it indirectly appears in the definition of
$\eR_2$.

\subsection*{Remarks}

1.~ A comprehensive discussion of the Poincar\'{e} group in
physics can be found in \cite{barut, bog}. Useful expressions for
Wigner rotations and their applications for massive particles are
given in \cite{hal,soo,har}.

\medskip
\noindent 2.~  In this transformation I do not assume any
additional normalization factors. A condition of unitarity is
$UU^\dag=U^\dag U=\1,$ but there also other conventions in the
literature.
\medskip

\noindent  3.~ A double infinity of the positive energy solutions of the
Dirac equation (functions $u^{(1/2)}_p$ and $u^{(-1/2)}_p$) spans
an improper basis of this space.  There is a one-to-one
correspondence between Wigner and Dirac wave functions. Basis
vectors of  Wigner and Dirac Hilbert spaces are in the one-to-one
correspondence \cite{bog},
\beq
u^{(1/2)}_p\Leftrightarrow|\half,p\9,\qquad
u^{(-1/2)}_p\Leftrightarrow|\half,p\9,
\eeq
while the wave functions are related by
\bea
& &\Psi^\alpha(p)=\psi_{1/2}(p)u^{(1/2)\alpha}_p +
\psi_{-1/2}(p)u_p^{(-1/2)\alpha} \label{corr}
 \\
& &
2m\psi_\sigma(p)=\sum_{\alpha=1}^4\overline{u}^{(-\sigma)}_{\alpha
p}\Psi^{\alpha}(p)
\eea

\medskip

\noindent 4.~ Another approach to the construction of the Wigner rotation
$\hat{D}$ is based on the homomorphism between Lorentz group and
$SL(2)$ \cite{bog}.
\medskip

\noindent 5.~ When not restricted to a single-particle space the
Wigner spin operator is given by
\beq
\bS=\half\sum_{\eta,\zeta}\mbox{\boldmath
$\sigma$}_{\eta\zeta}\int\!d\mu(p) (\hat{a}^\dag_{\eta
p}\hat{a}_{\zeta p}+\hat{b}^\dag_{\eta p}
\hat{b}_{\sigma p}),
\eeq
where $\hat{a}^\dag_{\eta p}$ creates a mode with a momentum $p$
and spin $\eta$ along the $z$-axis, etc. A comparison of different
spin operators can be found in \cite{ter3}.
\medskip

\noindent 6.~
If one works with the 4-vectors, then in the helicity gauge the
polarization vector is given by $\epsilon_p=(0,\bep_{\bp})$. A
formal connection between helicity states and polarization vectors
is made by first observing
 that  three spin-1 basis states can be constructed
from the components of a symmetric spinor of rank 2. Unitary
transformations of this spinor that are induced by $\eR$ are in
one-to-one correspondence with transformations by $\eR$ of certain
linear combinations of a spatial vector. In particular,
transformations of the helicity $\pm 1$ states induced by
rotations are equivalent to the rotations of $(\bep_{k_S}^1\pm
i\bep_{k_S}^2)/\sqrt{2}$ (the $z$-axis is the initial quantization
direction). While $p_\mu \epsilon_p^\mu=0$ gauge condition is
Lorentz-invariant, the spatial orthogonality is not.  The role of
gauge transformations in preserving the helicity gauge and some
useful expressions for the phase that photons acquire can be found
in
\cite{am,lpt,bga}

\section{Implications of quantum Lorentz transformations}
\subsection{Reduced density matrices}

In a relativistic system whatever is outside the past light cone
of the observer is unknown to him, but also cannot affect his
system, therefore does not lead to decoherence (here, I assume
that no particle emitted by from the outside the past cone
penetrates into the future cone). Since different observers have
different past light cones, by tracing out they exclude from their
descriptions different parts of spacetime. Therefore any
transformation law between them must tacitly assume that the part
excluded by one observer is irrelevant to the system of another.

Another consequence of relativity is that there is a hierarchy of
dynamical variables: {\it primary variables\/} have relativistic
transformation laws that depend only on the Lorentz transformation
matrix $\Lambda$ that acts on the spacetime coordinates. For
example, momentum components are primary variables. On the other
hand, {\it secondary variables\/} such as spin and polarization
have transformation laws that depend not only on $\Lambda$, but
also on the momentum of the particle. As a consequence, the
reduced density matrix for secondary variables, which may be well
defined in any coordinate system, has no transformation law
relating its values in different Lorentz frames.

Moreover, an unambiguous definition of the reduced density matrix
 is possible only if the secondary
degrees of freedom are unconstrained, and photons are the simplest
example when this definition fails. In the absence of a general
prescription, a case-by-case treatment is required. I describe a
particular construction, valid with respect to a certain class of
tests.

\subsection{Massive particles}

For a massive qubit  the usual definition of quantum entropy has
no invariant meaning. The reason is that under a Lorentz boost,
the spin undergoes a Wigner rotation, that as shown on Fig.~1 is
controlled both by the classical data and the corresponding
momentum. Even if the initial state is a direct product of a
function of momentum and a function of spin, the transformed state
is not a direct product. Spin and momentum become entangled.

Let us define a reduced density matrix,
\beq
\rho=\int d\mu(p)\psi(p)\psi^\dagger(p).
\eeq
It gives statistical predictions for the results of measurements
of spin components by an ideal apparatus which is not affected by
the momentum of the particle. Note that I tacitly assumed that the
relevant observable is the Wigner spin. The spin entropy is
\beq
S=-\tr(\rho\log\rho)=-\sum\lambda_j\log\lambda_j,
\eeq
where $\lambda_j$ are the eigenvalues of $\rho$.

As usual, ignoring some degrees of freedom leaves the others in a
mixed state. What is not obvious is that in the present case the
amount of mixing depends on the Lorentz frame used by the
observer.  Indeed consider another observer (Bob) who moves with a
constant velocity with respect to Alice who prepared that state.
In the Lorentz frame where Bob is at rest, the state is given by
Eq.~(\ref{state}).

\begin{figure}[htbp]
\epsfxsize=0.55\textwidth
\centerline{\epsffile{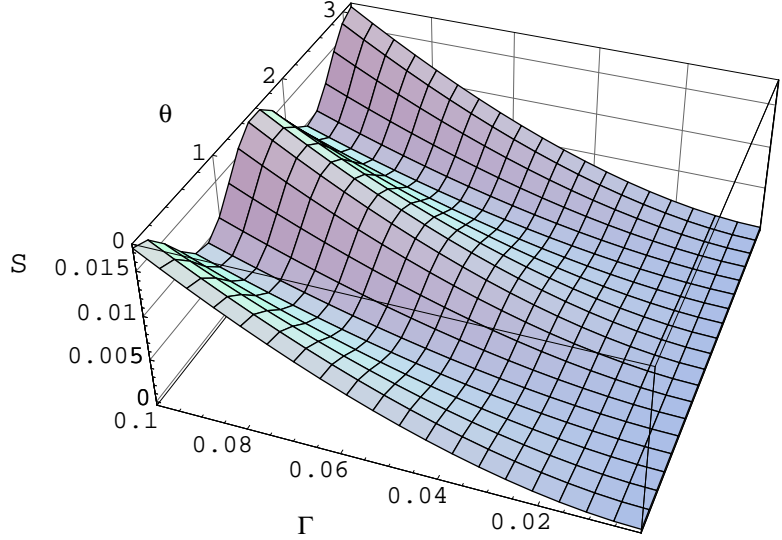}} \vspace*{-0.1cm} \caption{\small{Dependence
of the spin entropy $S$, in Bob's frame, on the values of the
angle $\theta$ and a parameter
$\Gamma=[1-(1-v^2)^{1/2}]\Delta/mv$.}}
\label{spinent}
\end{figure}

As an example, take a particle prepared by Alice to be
\beq
 |\Psi\9=\chi\int \psi(p)|p\9d\mu(p)\label{rest},\qquad \chi={\zeta\choose\eta}
\eeq
where $\psi$ is concentrated near zero momentum and has a
characteristic spread $\Delta$. Spin density matrices of all the
states that are given by Eq.~(\ref{rest}) are
\beq
\rho={|\zeta|^2~~\zeta\eta^*\choose \zeta^*\eta~~ |\eta|^2},
\eeq
and are independent of the specific form of $\psi(p)$.  To make
calculations explicit (and simpler) I take the wave function to be
Gaussian, $\psi(p)=N\exp(\bp^2/2\Delta^2)$, where $N$ is a
normalization factor. Spin and momentum are not entangled, and the
spin entropy is zero. When that particle is described in Bob's
Lorentz frame, moving with velocity $v$  at the angle $\theta$
with Alice's $z$-axis, a detailed calculation shows that the the
spin entropy is positive
\cite{rmp}. This phenomenon is illustrated in Fig.~\ref{spinent}.
A relevant parameter, apart from the angle $\theta$, is in the
leading order in momentum spread,
\beq
\Gamma=\frac{\Delta}{m}\,\frac{1-\sqrt{1-v^2}}{v},
\eeq
where $\Delta$ is the momentum spread in Alice's frame.  The
entropy has no invariant meaning, because the reduced density
matrix $\tau$ has no covariant transformation law, except in the
limiting case of sharp momenta.  Only the complete density matrix
transforms covariantly.

I outline some of the steps in this derivation. First, we
calculate the rotation parameters $(\omega,
\hat{\mathbf{n}})$ of the orthogonal matrix $W(\Lambda,p)$ for a
general momentum. The rotation axis and angle are given by
\beq
\mathbf{\hat{n}=\hat{\bv}\times\hat{p}},\qquad
\cos\theta=\mathbf{\hat{\bv}\cdot\hat{p}},\qquad 0\leq\theta\leq\pi
\eeq
where $\hat{\bv}$ is  boost's direction, while the leading order
term for the angle is
\beq
\omega  =
\frac{1-\sqrt{1-v^2}}{v}\frac{p}{m}\sin\theta
-\eO\left(\frac{p^2}{m^2}\right)
\label{omeg}
\eeq

Without a loss of generality we can make another simplification.
We can choose our coordinate frame in such a way that  both
$\zeta$ and $\eta$ are real. The matrix
$D[W(\Lambda,\Lambda\!^{-1}p)]$ takes the form
\beq
D[W(\Lambda,p')]=\sigma_0\cos{\omega\over 2}-i\sin{\omega\over
2}(-\sin\phi \,\sigma_x+\cos\!\phi\,\sigma_y),
\eeq
where $(\theta,\phi)$ are the spherical angle of  $\bp'$ (to be
consistent with Eq.~(\ref{state}) momentum in Alice frame carries
a prime, $p'=\Lambda^{-1}p$). The reduced density matrix in Bob's
frame is
\beq
\rho^B_{\sigma\xi}=\int \!d\mu(p)D_{\sigma\nu}D^*_{\xi\lambda}
\psi_\nu(p')\psi^*_\lambda(p').
\eeq
The  symmetry of $\psi(\Lambda^{-1}p)$ is cylindrical. Hence the
partial trace is taken by performing  a momentum integration in
cylindrical coordinates.  This simplification is a result of the
spherical symmetry of the original $\psi$. The two remaining
integrations are performed by first expanding in powers of
$p/\Delta$ and taking Gaussian integrals. Finally,
\beq
\rho'=\left(
\ba{cc} \zeta^2(1-\Gamma^2/4)+\eta^2\Gamma^2/4 & \zeta\eta^*(1-\Gamma^2/4) \\
\zeta^*\eta(1-\Gamma^2/4) & \zeta^2\Gamma^2/4+\eta^2(1-\Gamma^2/4)
\ea \right).
\eeq

Fidelity can be used to estimate the difference between the two
density matrices. It is defined as
\beq
f=\chi^\dag\rho'\chi,
\eeq
and it is easy to get an analytical result for this quantity. Set
$\zeta=\cos\theta$ and $\eta=\sin\theta$. Then
\beq
f=1-\frac{\Gamma^2}{2}\left(3+\frac{\cos{4\theta}}{8}\right).
\eeq

Consider now a pair of orthogonal states that were prepared by
Alice, e.g. the above state with $\chi_1=(1,0)$ and
$\chi_2=(0,1)$. How well can moving Bob distinguish them? I use
the simplest criterion, namely the probability of error $P_E$,
defined as follows: an observer receives a single copy of one of
the two known states and performs any operation permitted by
quantum theory in order to decide which state was supplied. The
probability of a wrong answer for an optimal measurement is
\beq
P_E(\rho_1,\rho_2)=\half-\mbox{$1\over4$}\,{\rm tr}
 \sqrt{(\rho_1-\rho_2)^2}.  \label{pe}
 \eeq
In Alice's frame $P_E=0$.  In Bob's frame the reduced density
matrices are
\beq
\rho^B_1=\left(\ba{cc}
1-\Gamma^2/4 & 0 \\
0 & \Gamma^2/4
\ea\right), \qquad \rho^B_2=\left(\ba{cc}
\Gamma^2/4 & 0 \\
0 & 1-\Gamma^2/4,
\ea\right)
\eeq
respectively. Hence the probability of error  is
$P_E(\rho_1,\rho_2)=\Gamma^2/4$.

\subsection{Photons}\label{photons}

The relativistic effects in photons are essentially different from
those for massive particles that were discussed above. This is
because photons have only two linearly independent polarization
states. As we know, polarization is a {\it secondary variable\/}:
states that correspond to different momenta belong to distinct
Hilbert spaces and cannot be superposed (an expression such as
$|\bep_\bk^\pm\9+|\bep_\bq^\pm\9$ is meaningless if $\bk\neq\bq$).
The complete basis $|p,\bep_\bp^\pm\9$  does not violate this
superselection rule, owing to the orthogonality of the momentum
basis. The reduced density matrix, according to the usual rules,
should be
\beq
\rho=\int\! d\mu(p)|\psi(p)|^2|p,\bal(\bp)\9\6p,\bal(\bp)|.
\label{rhost}
\eeq
However, since $\xi$ in Eq.~(\ref{dss}) depends on the photon's
momentum even for ordinary rotations, this object will have no
transformation law at all. It is still possible  define an
``effective''
 density matrix adapted to a specific method of measuring
polarization \cite{pt03,aw}. I describe one  such scheme.

The labelling of polarization states by Euclidean vectors
$\bep_\bp^\pm$ suggests the use of a $3\times 3$ matrix with
entries labelled $x$, $y$ and $z$. Classically, they correspond to
different directions of the electric field. For example, a
component $\rho_{xx}$ would give the expectation values of
operators representing the polarization in the $x$ direction,
seemingly irrespective of the particle's momentum.

To have a momentum-independent polarization is to admit
longitudinal photons. Momentum-independent polarization states
thus consist of physical (transverse) and unphysical
(longitudinal) parts, the latter corresponding to a polarization
vector $\bep^\ell=\hbk$. For example, a generalized polarization
state along the $x$-axis is
\beq
|\hbx\9=x_+(\bp)|\bep^+_\bp\9+x_-(\bp)|\bep^-_\bp\9+
x_\ell(\bp)|\bep^\ell_\bp\9,\label{decomp}
\eeq
where $x_\pm(\bp)=\hbx\cdot\bep^\pm_\bp$, and $x_\ell(\bp)=
\hbx\cdot\hbk=\sin\theta\cos\phi$. It follows that
$|x_+|^2+|x_-|^2+|x_\ell|^2=1$, and we thus define
\beq
\be_x(\bp)=\frac{x_+(\bp)\bep^+_\bp+x_-(\bp)\bep^-_\bp}
{\sqrt{x_+^2+x_-^2}} \label{physdir},
\eeq
as the polarization vector associated with the $x$ direction. It
follows from (\ref{decomp}) that $\6\hbx|\hbx\9=1$ and
$\6\hbx|\hby\9=\hbx\cdot\hby=0$, and likewise for the other
directions, so that
\beq
|\hbx\9\6\hbx|+|\hby\9\6\hby|+|\hbz\9\6\hbz|=\1_p,\label{xyz}
\eeq
where $\1_p$ is the unit operator in momentum space.

To the direction $\hbx$ there corresponds a projection operator
\beq
P_{xx}=|\hbx\9\6\hbx|\otimes \1_p=|\hbx\9\6\hbx|\otimes \int
d\mu(k)|\bp\9\6\bp|,
\eeq
 The action of
$P_{xx}$ on $|\Psi\9$ follows from Eq.~(\ref{decomp}) and
$\6\bep^\pm_\bp|\bep^\ell_\bp\9=0$. Only the transverse part of
$|\hbx\9$ appears in the expectation value:
\beq
\6\Psi|P_{xx}|\Psi\9=\int d\mu(p)|\psi(p)|^2|x_
 +(\bp)\alpha_+^*(\bp)+x_-(\bp)\alpha_-^*(\bp)|^2.
 \eeq
It is convenient to write the transverse part of $|\hbx\9$ as
\beq
|\bb_x(\bp)\9  \equiv
 (|\bep^+_\bp\9\6\bep^+_\bp|+|\bep^-_\bp\9\6\bep^-_\bp|)|\hbx\9
 = x_+(\bp)|\bep^+_\bp\9+x_-(\bp)|\bep^-_\bp\9.
\label{vector}
\eeq
Likewise define  $|\bb_y(\bp)\9$ and $|\bb_z(\bp)\9$. These three
state vectors are neither of unit length nor mutually orthogonal.

Finally, a POVM element $E_{xx}$ which is the physical part of
$P_{xx}$, namely is equivalent to $P_{xx}$ for physical states
(without longitudinal photons) is
\beq
E_{xx}=\int d\mu(k)|p,\bb_x(\bp)\9\6p,\bb_x(\bp)|,
\eeq
and likewise for the other directions. The operators $E_{xx}$,
$E_{yy}$ and $E_{zz}$ indeed form a POVM in the space of physical
states, owing to Eq.~(\ref{xyz}).

To complete the construction of the density matrix, we introduce
additional directions. Following the standard practice of state
reconstruction, we consider $P_{x+z,x+z}$, $P_{x+iz,x+iz}$ and
similar combinations. For example,
\beq
P_{x+z,x+z}=\half(|\hbx\9+|\hbz\9)(\6\hbx|+\6\hbz|)\otimes\1_p.
\eeq
The diagonal elements of the new polarization density matrix are
defined as
\beq
\rho_{mm}=\6 \Psi|E_{mm}|\Psi\9,\qquad m=x,y,z,
\eeq
and the off-diagonal elements are recovered by combinations such
as
\beq
\rho_{xz}=\6\Psi(|\hbx\9\6\hbz|\otimes\1_p)|\Psi\9=
\6\Psi|E_{x+z,x+z}-iE_{x-iz,x-iz}+(1-i)(E_{xx}-E_{zz})/2|\Psi\9.
\eeq
Denote $|\hbx\9\6\hbz|\otimes\1_p$ as $P_{xz}$, and its
``physical" part by $E_{xz}$. Then the effective polarization
density matrix is
\beq
\rho_{mn}=\6\Psi|E_{mn}|\Psi\9=
\int d\mu(k)|f(\bp)|^2\6\bal(\bp)|\bb_m(\bp)\9\6\bb_n(\bp)|\bal(\bp)\9
, \quad m,n,=x,y,z. \label{reduced}
\eeq
It is interesting to note that this derivation gives a direct
physical meaning to the naive definition of a reduced density
matrix,
\beq
\rho^{\rm naive}_{mn}=\int d\mu(p)|\phi(p)|^2\bal_m(\bp)\bal_n^*(\bp)
=\rho_{mn}
\eeq
It is possible to show that this POVM actually corresponds to a
simple photodetection model \cite{lt}.

The basis states $|p,\bep_\bp\9$ are direct products of momentum
and polarization. Owing to the transversality requirement
$\bep_\bp\cdot\bp=0$, they remain direct products under Lorentz
transformations. All the other states have their polarization and
momentum degrees of freedom entangled. As a result, if one is
restricted to polarization measurements as described by the above
POVM, {\it there do not exist two orthogonal polarization
states\/}. In general, any measurement procedure with finite
momentum sensitivity will lead to the errors in identification, as
demonstrated as follows

Let two states $|\Phi\9$ and $|\Psi\9$  be two orthogonal
single-photon states. Their reduced polarization density matrices,
$\rho_\Phi$ and $\rho_\Psi$, respectively, are calculated using
Eq.~(\ref{reduced}). Since the states are entangled, the von
Neumann entropies of the reduced density matrices, $S=-{\rm
tr}(\rho\ln\rho)$, are positive. Therefore, both matrices are at
least of rank two. Since the overall dimension is~3, it follows
that ${\rm tr}(\rho_\Phi\rho_\Psi)>0$ and these states are not
perfectly distinguishable. An immediate corollary is that photon
polarization states cannot be cloned perfectly, because the
no-cloning theorem  forbids exact copying of unknown
non-orthogonal states.

In general, any measurement procedure with finite momentum
sensitivity will lead to the errors in identification. First I
present some general considerations and then illustrate them with
a simple example. Let us take the $z$-axis to coincide with the
average direction of propagation so that the mean photon momentum
is $k_A\hbz$. Typically, the spread in momentum is small, but not
necessarily equal in all directions. Usually the intensity profile
of laser beams has cylindrical symmetry, and we may assume that
$\Delta_x\sim\Delta_y\sim\Delta_r$ where the index $r$ means
radial. We may also assume that $\Delta_r{\gg}\Delta_z$. We then
have
\beq
f(p)\propto f_1[(p_z-k_A)/\Delta_z]\,f_2(p_r/\Delta_r).
\eeq
 We  approximate
\beq
\theta\approx\tan\theta\equiv p_r/p_z\approx p_r/k_A.
\label{theta}
\eeq
In  pictorial language, polarization planes for different momenta
are tilted by angles up to $\sim\Delta_r/k_A$, so that we expect
an error probability of the order $\Delta_r^2/k_A^2$. In the
density matrix $\rho_{mn}$ all the elements of the form
$\rho_{mz}$ should vanish when $\Delta_r\to0$. Moreover, if
$\Delta_z\to0$, the non-vanishing $xy$ block goes to the usual
(monochromatic) polarization density matrix,
\beq
\rho_{{\rm pure}}=\left(\bay{ccc}
|\alpha|^2 & \beta & 0\\
\beta^* & 1-|\alpha|^2 &0\\
0 & 0 &0
\eay\right).
\eeq

As an example,  consider two states which, if the momentum spread
could be ignored, would be $|k_A\hat{\bf z},\bep^\pm_{k_A\hat{\bf
z}}\9$. To simplify the calculations we assume a Gaussian
distribution:
\beq
f(p)=Ne^{-(p_z-k_A)^2/2\Delta_z^2}e^{-p_r^2/2\Delta_r^2},
\eeq
where $N$ is a normalization factor and $\Delta_z, \Delta_r\ll
k_A$. In general the spread in $p_z$ may introduce an additional
incoherence into  density matrices, in addition to the effect
caused by the transversal spread. However, when all momentum
components carry the same helicity, this spread results in
corrections of
 the higher order. In the example below we take the polarization
components to be $\bep^\pm_\bp\equiv R(\hbk)\bep^\pm_{k_S}$. That
means we have to analyze  the states
\beq
|\Psi_\pm\9=\int d\mu(p) f(p)|p,\bep^\pm_\bp\9\label{st},
\eeq
where $f(p)$ is given above.

It is enough to expand $R(\hbk)$ up to second order in $\theta$.
The reduced density matrices are calculated by techniques similar
to those for massive particles, using rotational symmetry around
the $z$-axis and normalization requirements. At the leading order
in $\Omega\equiv\Delta_r/k_A$
\beq
\rho_+=\half(1-\half\Omega^2)\left(\bay{ccc}
1 & -i & 0\\
i & 1 &0\\
0 & 0 &0
\eay\right)+\half\Omega^2\left(\bay{ccc}
0 & 0 & 0\\
0 & 0 & 0\\
0 & 0 & 1
\eay\right),
\eeq
and $\rho_-=\rho_+^*$. At the same level of precision,
\beq
P_E(\rho_+,\rho_-)=\Delta_r^2/4k_A^2.
\eeq
It is interesting to note that the optimal strategy for
distinguishing between these two states is a polarization
measurement in the $xy$-plane. Then the effective $2\times 2$
density matrices are perfectly distinguishable, but there is a
probability $\Omega^2/2$ that no photon will be detected at all.
The above result is valid due to the special form of the states
that we had chosen. Potential errors in the upper $2\times 2$
blocks are averaged out in the integration over~$\phi$. The effect
becomes important  when, e.g., a plain monochromatic wave
undergoes a strong focusing. Then $\Omega\approx l/f$, where $l$
is the aperture radius and $f$ is the focal length \cite{lt}.

Now let us turn to the distinguishability problem from the point
of view of a moving observer, Bob.  The probability of an error is
still given by Eq.~(\ref{pe}), but the parameters depend on the
observer's motion. Assume again  that Bob moves along the $z$-axis
with a velocity $v$.   To calculate Bob's reduced density matrix,
we must transform the complete state, and  then take a partial
trace.

Reduced density matrices of $|\Psi_\pm\9$ in both frames are given
by the expression
\beq
(\rho_\pm)^{A,B}_{mn}=\int d\mu(p)|f(p)^{A,B}|^2\6
R(\hbk)\bep_{k_S}^\pm|\bb_m(\bp)\9\6\bb_n(\bp)|R(\hbk)\bep_{k_S}^\pm\9.
\eeq
This is due to the following two reasons. First, $|\bb_m(\bp)\9$
are defined by Eq.~(\ref{vector})  in any frame, while pure boosts
preserve the orientation of the coordinate axes in 3-space, and
therefore do not affect the indices of $\rho_{mn}$. Second, phases
acquired by polarization states  cancel out, since we choose the
states $|\bp,\bal(\bp)\9$ to be the helicity eigenstates.

 Calculation of Bob's density matrix is similar to the previous cases.
 The only
frame dependent expression in (\theequation) is
$f^B(p)=f^A(\Lambda^{-1}p)$.  A boost along the $z$-axis preserves
$k_r$ and $\phi$. On the other hand,
\beq
k^B_z \approx k_A\sqrt{\frac{1-v}{1+v}}.
\eeq
Since everything else in the integral remains the same, the effect
of relative motion is given by a substitution
\beq
\Omega^B=\sqrt{\frac{1+v}{1-v}}\Omega^A, \qquad
P^B_E=\frac{1+v}{1-v}P^A_E, \label{dope}
\eeq
so Bob can distinguish the signals either better or worse than
Alice \cite{pt03}.

\subsection*{Remarks}

\noindent 1. A modification of the spin operator \cite{cw} will allow for a
momentum-independent transformation of the spin density matrix
between two frames that are related by a fixed Lorentz
transformation $\Lambda_{12}$. Its relation to our scheme is
discussed in \cite{czcom}.
\medskip

\noindent 2. An  additional motivation for introduction of effective polarization
density matrices cames from the analysis  of one-photon scattering
\cite{aw}.
\medskip

\noindent 3. I have discussed only discrete variables. To explore the relativistic
effects with continuous variables \cite{ralph} it is convenient to
express the quantum Lorentz transformations in terms of mode
creation and annihilation operators \cite{kokbrau}.

\section{Communication channels}
What happens when Alice and Bob that are in a relative motion try
to communicate? Assume that they use qubits that were described
above. Under a general Lorentz transformation $\Lambda$ that
relates Alice's and Bob's frames, the state of this qubit will be
transformed due to three distinct effects, which are:

 (i) A Wigner rotation due to the Lorentz boost $\Lambda$, which occurs
even for momentum eigenstates. If $\Lambda$ is known, then to the
extent that the wave-packet spread can be ignored, this is
inconsequencial.

(ii) A decoherence due to the entangling of spin and momentum
under the Lorentz transformation $\Lambda$ because the particle is
not in a momentum eigenstate. Although reduced or effective
density matrices have no general transformation rule,  such rules
can be established for particular classes of experimental
procedures. We can then ask how these effective transformation
rules, $\rho'=T(\rho)$, fit into the framework of general state
transformations. E.g., for the massive qubit of Sec.~4.2 the
effective transformation is given by
\beq
\rho'=\rho(1-\frac{\Gamma^2}{4})+(\sigma_x\rho\sigma_x+\sigma_y\rho\sigma_y)
\frac{\Gamma^2}{8}. \label{chanex}
\eeq
If $\Lambda$ is known and it is possible to implement the
operators that were mentioned in the Remark~1 above, then this
effect is absent. Otherwise, this noise is unavoidable. Still, it
is worth to keep in mind that the motion can improve the message
fidelity, as in Eq.~(\ref{dope}).

(iii) Another kind of decoherence arises due to Bob's lack of
knowledge about the transformation relating his reference frame to
Alice's frame. Using the techniques of the decoherence-free
subspaces, it is possible to eliminate this effect completely. E.
g., for massive particles four physical qubits may be used to
encode a logical qubit, while for photons $2\rightarrow 1$
encoding is sufficient. In both cases using the block encoding it
is possible to reach an asymptotically unit efficiency \cite{bt}.

Entanglement between the ``qubit'' and spatial degrees of freedom
leads to an interesting complication of the analysis. It is known
\cite{rmp}, that the dynamics of a subsystem my be not completely
positive, there is a prior entanglement with another system and
the  dynamics is not factorizable. Since in Eq.~(\ref{dope}) and
in the discussion following Eq.~(\ref{pe}) we have seen that
distinguishability {\it can\/} be improved, we conclude that these
transformations are {\it not completely positive\/}. The reason is
that the Lorentz transformation acts not only on the
``interesting'' discrete variables, but also on the primary
momentum variables that we elected to ignore and to trace out, and
its action on the interesting degrees of freedom depends on the
``hidden'' primary ones. Of course, the complete state, with all
the variables, transforms unitarily and distinguishability is
preserved.

This technicality has one important consequence. In quantum
information theory quantum channels are described by completely
positive maps that act on qubit states. Qubits themselves are
realized as discrete degrees of freedom of various particles. If
relativistic motion is important, then not only does the vacuum
behave as a noisy quantum channel, but the very representation of
a channel by a CP map fails.

\section{Entanglement and different Lorentz observers}\label{entrel}

In this section I consider only two-particle states. Even in this
simple setting there are several possible answers to the question
what happens to the entanglement, depending on the details of the
question. Since the quantum Lorentz transformation is given by a
tensor product  $U_1(\Lambda)\otimes U_2(\Lambda)$, the overall
entanglement between the states is Lorentz-invariant.

Let us assume that the states can be approximated by momentum
eigenstates. Then, the same conclusion applies to the spin-spin
(or polarization-polarization) entanglement between the particles,
and it is possible to write an appropriate entanglement measures
that capture the effects of particle statistics and
Lorentz-invariance of the entanglement \cite{soo}. However, it
does not mean that this invariance will be observed in an
experiment, or that the violation of Bell-type inequalities that
is observed in an experiment that is performed in Alice's frame
will be observed if the same equipment is placed  in Bob's frame.

While a field-theoretical analysis shows that violations of
Bell-type inequalities are generic, there are conditions that are
imposed on the experimental procedures that are used to detect
them. Consider  the CHSH inequality. For any two spacelike
separated regions and any pairs of of operators, $a$, $b$, there
is a state $\rho$ such that the CHSH inequality is violated, i.e.,
$\zeta(a,b,\rho)>1$. With additional technical assumptions the
existence of a maximally violating state $\rho_m$ can be proved:
\beq
\zeta(a,b,\rho_m)=\sqrt{2} \label{violall},
\eeq
for any spacelike separated regions $\cO_L$  and $\cO_R$. It
follows from convexity arguments that states that maximally
violate Bell inequalities are pure. What are then the operators
that lead to the maximal violation? It was shown \cite{sw} that
the operators $A_j$ and $B_k$ that give $\zeta=\sqrt{2}$ satisfy
$A_j^2=\1$ and $A_1A_2+A_2A_1=0$, and likewise for $B_k$. If we
define $A_3:=-i[A_1,A_2]/2$, then these three operators have the
same algebra as Pauli spin matrices.

In principle the vacuum state may lead to the maximal violation of
Bell-type inequalities. Their observability was discussed in
\cite{vac}.

 The operators \cite{c97}
\beq
A_i=2\left[\frac{m}{p^0}{\boa}_i+
\left(1-\frac{m}{p^0}\right)(\boa\cdot\bn)\bn\right]\cdot\bS
\equiv2\mbox{\boldmath $\alpha$}(\boa,\bp)\cdot\bS,
\eeq
where $\bS$ is the Wigner spin operator and $\bn=\bp/|\bp|$ appear
quite naturally as the candidates for the  measurement
description. The length of the auxiliary vector $\mbox{\boldmath
$\alpha$}$ is
\beq
|\mbox{\boldmath$\alpha$}|=
\frac{\sqrt{(\bp\cdot\boa)^2+m^2}}{p^0},
\eeq
so  generically $A_i^2=\mbox{\boldmath $\alpha$}^2\1<\1$, and
indeed, the degree of violation decreases with the velocity of the
observer. Nevertheless, it is always possible to compensate for a
Wigner rotation by an appropriate choice of the operators
\cite{rmp}.

Realistic situations involve wave packets. For example, a general
spin-$\half$  two-particle state  may be written as
\beq
|\Upsilon_{12}\9=\sum_{\sigma_1,\sigma_2}\int
d\mu(p_1)d\mu(p_2)g(\sigma_1\sigma_2,\bp_1,\bp_2)|\bp_1,\sigma_1
;\bp_2,\sigma_2\9.
 \label{entfig}
\eeq
For particles with well defined momenta, $g$  sharply peaks at
some values $\bp_{10}$, $\bp_{20}$. Again, a boost to any Lorentz
frame $S'$ will result in a unitary $U(\Lambda)\otimes
U(\Lambda)$, acting on each particle separately, thus preserving
the entanglement. Nevertheless, since  they can change
entanglement between different degrees of freedom of a single
particle, the spin-spin entanglement is frame-dependent as well.
Having investigated the reduced density matrix for
$|\Upsilon_{12}\9$ and made explicit calculations for the case
where $g$ is a Gaussian, as in the Sec.~4.2 above, it is possible
to show  that if two particles are maximally entangled in a common
(approximate) rest frame (Alice's frame), then the concurrence, as
seen by a Lorentz-boosted Bob, decreases when $v\to 1$. Of course,
the inverse transformation from Bob to Alice will increase the
concurrence \cite{ga}.  Thus, we see that that spin-spin
entanglement is not a Lorentz invariant quantity, exactly as spin
entropy is not a Lorentz scalar. Relativistic properties of the
polarization entanglement are even more interesting \cite{bga},
since there is no frame where polarization and momentum are
unentangled.

\section{The omissions \& perspectives}
Because of the lack of space I am only going to mention the
various fascinating areas of the interplay between quantum
information theory and relativistic physics. Quantum field theory
provides us with new situations that should be investigated. For
example, it is possible to ask all the usual questions about
entanglement, distillability, etc and their invariance\cite{rmp,
vw, dt04,eis}. So far we discussed only observers that move with
constant velocity. An accelerated observer sees Unruh radiation.
It leads to a host of interesting effects if we consider a
teleportation between a stationary and accelerated observer
\cite{am03,ivette,unruh}. Dynamical entanglement --- the one appearing in the
scattering processes or between the decay products also have been
investigated
\cite{har,man04,solano}.

Going to more exotic settings, I just mention that black hole
physics, cosmology, loop quantum gravity and string theory provide
extremely interesting scenarios where the questions of information
can and should be asked\cite{rmp,astro,flo, lit, lit2, dt05,my05}


\begin{thebibliography}{99}
\bibitem{rmp} A. Peres and D. R. Terno, \textit{Rev. Mod. Phys.} {\bf 76} (2004), 93.
\bibitem{nc} M. A. Nielsen and I. L. Chuang, {\it
Quantum Computation and Quantum Information\/}, Cambridge
University Press, New York, 2000.
\bibitem{keyl} M. Keyl, Phys. Rep. {\bf 369} (2002), 431.
\bibitem{bgl} P. Busch,  M. Grabowski, and P. J. Lahti,  {\it
Operational Quantum Physics\/}, Springer, Berlin, 1995.

\bibitem{rm1} H.-P. Breuer and F. Petruccione (eds.), \textit{
Relativistic Quantum Measurement and Decoherence}, Springer,
Berlin, 2000.
\bibitem{rm2} H.-P. Breuer (ed.), \textit{
Relativistic Quantum Measurement}, Springer, Berlin, 2000.

\bibitem{ben99} C. H. Bennett, D. P. DiVincenzo, C. A. Fuchs, T. Mor, E. Rains,
P. W. Shor, J. A. Smolin, and W. K. Wootters, \textit{Phys. Rev.
A}  {\bf59} (1999), 1070.
\bibitem{bek01} D. Beckman, D. Gottesman, M. A. Nielsen, and J.
Preskill, \textit{Phys. Rev. A} {\bf64} (2001), 052309.
\bibitem{egg02} T. Eggeling,  Schlingemann, and R. F. Werner,
\textit{ Europhys. Lett.} {\bf57} (2002), 782.
\bibitem{gr02} B. Groisman and B. Reznik,\textit{ Phys. Rev. A} {\bf66}
(2002), 022110.
\bibitem{vai03} L. Vaidman, \textit{Phys. Rev. Lett.} {\bf 90}  (2003) 010402.
\bibitem{sf} M. Florig and S. J. Summers,  \textit{J. Math. Phys.} {\bf38} (1997),
1318.
\bibitem{k99} A. Kent, \textit{Phys. Rev. Lett.} {\bf 83} (1999), 1447.
\bibitem{k03} A.  Kent, \textit{Phys. Rev. Lett.} {\bf 90} (2003), 237901.

\bibitem{keller} O. Keller, \textit{Phys. Rep.\/} {\bf 411}
(2005), 1.
\bibitem{tol} M. Toller, \textit{Phys. Rev. A} {\bf 59} (1999),
960.
\bibitem{maz} S. Mazzucchi, \textit{J. Math. Phys.} {\bf 42} (2001),
2477.
\bibitem{wkt} W.-K. Tung, \textit{Group Theory in Physics\/}, World
Scientific, Singapore, 1985.
\bibitem{wei} S. Weinberg, {\it The Quantum Theory of Fields\/}
  Vol.~1, Cambridge University, Cambridge, 1995.

\bibitem{barut} A. O. Barut and R. Raczka, \textit{Theory of Group Representations and
 Applications\/}, World Scientific, Singapore, 1987.
 \bibitem{bog}  N.N. Bogolubov,  A. A. Logunov, A. I. Oksak, and I. T. Todorov,
 {\it General Principles of Quantum Field Theory\/}, Kluwer,
Dordrecht, 1990.
\bibitem{hal}F. R. Halpern, {\it Special Relativity and
Quantum Mechanics\/}, Prentice-Hall, Englewood Cliffs, 1968.
\bibitem{soo} C. Soo and C. C. Y. Lin, \textit{Int. J. Quant. Info.}
{\bf 2} (2004), 183.
\bibitem{har} N. L. Harshman, \textit{Phys. Rev. A} {\bf 71}
(2005), 022312.
\bibitem{ter3} D. R. Terno, \textit{Phys. Rev. A}  {\bf 67} (2003),
014102.
\bibitem{am} P. M. Alsing  and G. J. Milburn, \textit{Quant. Info.
Comp.\/} {\bf2} (2002), 487.
\bibitem{lpt} N. H. Lindner,  A. Peres, and D. R. Terno, \textit{J. Phys. A} {\bf36} (2003),
L449.
\bibitem{bga} A. J. Bergou,  R. M. Gingrich, and C. Adami, \textit{Phys. Rev.
A\/} {\bf{68}} (2003), 042102.
\bibitem{pt03} A. Peres and D. R. Terno, \textit{J. Mod. Opt.\/}
{\bf 50} (2003), 1165.
\bibitem{aw} A. Aiello and J. P. Woerdman, \textit{Phys. Rev. A} {\bf 70}
(2004), 023808.
\bibitem{lt} N. H. Lindner and D. R. Terno, \textit{J. Mod. Opt.\/} {\bf 52} (2005), 1177.
\bibitem{cw} M. Czachor and M. Wilczewski, \textit{Phys. Rev. A} {\bf
68} (2003), 010302(R).
\bibitem{czcom} M. Czachor, \textit{Phys. Rev. Lett.} {\bf 94}
(2005),  078901.
\bibitem{ralph} P.  Kok, T. C. Ralph, and G. J.  Milburn, \textit{Quant. Info. Comp.} {\bf 5}
(2005), 239.
\bibitem{kokbrau} P. Kok and S. L. Braunstein, \textit{Int. J.
Quant. Info.} {\bf 4} (2006), in press.
\bibitem{bt}  S. D. Bartlett and D. R. Terno, \textit{Phys. Rev. A\/}  {\bf 71} (2005),
012302.
\bibitem{sw} S. J. Summers  and R. Werner, \textit{J. Math. Phys.\/} {\bf
28} (1987), 2440.
\bibitem{c97} M. Czachor, \textit{Phys. Rev. A} {\bf 55} (1997), 72.
\bibitem{vac} B. Reznik, A. Retzker and J. Silman, \textit{Phys. Rev. A } {\bf 71} (2005),
042104.
\bibitem{ga} R. M. Gingrich  and C. Adami, \textit{Phys. Rev. Lett.\/} {\bf 89} (2002),  270402.

\bibitem{vw}  R. Verch and R. F. Werner, \textit{Rev. Math. Phys.} {\bf 17} (2005), 545.



\bibitem{dt04} D. R. Terno, \textit{Phys. Rev. Lett.\/} {\bf 93} (2004)
051303.
\bibitem{eis} M. B. Plenio, J. Eisert, J. Dreissig, and M. Cramer \textit{ Phys. Rev. Lett.} {\bf 94}
(2005),
060503.
\bibitem{am03} P. M. Alsing  and G. J. Milburn, \textit{Phys. Rev. Lett.\/} {\bf91} (2003),
180404.
\bibitem{ivette}  I. Fuentes-Schuller and  R. B. Mann, \textit{Phys. Rev. Lett.} {\bf 95 }(2005),
120404.
\bibitem{unruh}  R. Sch\"utzhold and  W. G. Unruh, e-print
quant-ph/0506028.
\bibitem{man04}  E. B. Manoukian and N. Yongram, \textit{Eur. J.
Phys. D\/} {\bf31} (2004), 137.
\bibitem{solano}  L. Lamata, J. Leon, and  E. Solano, e-print
quant-ph/0509021.
\bibitem{astro} D. Campo and R. Parentani, e-print
astro-ph/0505376.
\bibitem{flo} O. Dreyer, F. Markopoulou, and L. Smolin, e-print
hep-th/0409056.
\bibitem{lit} E. R. Livine and D. R. Terno, \textit{Phys. Rev. A}
{\bf 72}  (2005),  022307.
\bibitem{lit2} E. R. Livine and D. R. Terno, e-print
gr-qc/0508085.
\bibitem{dt05} D. R. Terno, e-print gr-qc/0505068.
\bibitem{my05} E. B. Manoukian and N. Yongram, \textit{Mod. Phys. Lett. A} {\bf 20}
(2005),
623.

\end{thebibliography}
\end{document}